\pdfoutput=1
\documentclass{article}

\usepackage[nonatbib,preprint]{neurips_2020}
\usepackage[numbers]{natbib}

\usepackage[utf8]{inputenc} 
\usepackage[T1]{fontenc}    
\usepackage{hyperref}       
\usepackage{url}            
\usepackage{booktabs}       
\usepackage{amsfonts}       
\usepackage{nicefrac}       
\usepackage{microtype}      
\usepackage{xcolor}
\usepackage{xspace}
\usepackage{tabularx}
\usepackage{amsmath,bm,multicol,multirow}
\usepackage{cleveref}
\usepackage[pdftex]{graphicx}
\usepackage{array, makecell}
\usepackage[scientific-notation=true]{siunitx}
\usepackage{subcaption}
\usepackage{chngpage}

\usepackage{ulem}
\crefformat{section}{\S#2#1#3} 
\crefformat{subsection}{\S#2#1#3}
\crefformat{subsubsection}{\S#2#1#3}
\DeclareMathOperator*{\argmax}{argmax} 
\DeclareMathOperator*{\argmin}{argmin}

\title{\LARGE A Comparison of Discrete Latent Variable Models \\ for Speech Representation Learning}

\author{%
Henry Zhou$^{\bigtriangledown}$\thanks{Work done while at Facebook during a Facebook AI Residency.} \And 
Alexei Baevski$^{\bigtriangleup}$ \And 
Michael Auli$^{\bigtriangleup}$ \AND
$^{\bigtriangledown}$ University of Toronto \And
$^{\bigtriangleup}$ Facebook AI Research 
}

\begin{document}

\maketitle

\begin{abstract}
Neural latent variable models enable the discovery of interesting structure in speech audio data. 
This paper presents a comparison of two different approaches which are broadly based on predicting future time-steps or auto-encoding the input signal.
Our study compares the representations learned by vq-vae and vq-wav2vec in terms of sub-word unit discovery and phoneme recognition performance.
Results show that future time-step prediction with vq-wav2vec achieves better performance. 
The best system achieves an error rate of 13.22 on the ZeroSpeech 2019 ABX phoneme discrimination challenge.
\end{abstract}

\section{Introduction}

Speech contains structure that algorithms processing this data reason about to make good predictions.
In the classical supervised learning paradigm both representation learning and making good predictions based on these representations are intertwined. 
A major limitation of this approach is that it requires large amounts of labeled data. 
Because of this, there has been much recent interest in algorithms which learn good representations or latent structure of speech without supervision.

Latent variable models have been heavily studied for speech applications, including voice conversion ~\cite{nakashika2015}, speaker identification and verification~\cite{lai2019}.
Recently, discrete representations learning through quantization~\cite{hu2017learningdiscrete,oord2017vqvae} have been used on inherently continuous data such as images and speech.
If the latent structure is modeled as a discrete set of units, then it can be evaluated in terms of semantic meaning such as in the ZeroSpeech challenge~\cite{zerospeech2019}.

There are various methods for learning quantized latent features and in this study, we focus on two popular approaches:
quantized latent features can be learnt through autoencoding, which reconstructs the original signal, either the raw waveform or spectrogram features~\cite{oord2017vqvae,vqvae_wavenet2019}.
Another approach learns latent features through predicting representations of future time-steps~\cite{schneider2019wav2vec,baevski2019vqwav2vec,alex2019unsupervised,harwath2019learning}.

In this study, we are interested in the quality of the discrete representations learnt by these two methods. 
In particular, we perform pre-training with either vq-vae or vq-wav2vec and evaluate the resulting representations in terms of phoneme recognition. This enables us to evaluate whether the discrete representations are helpful in distinguishing phonemes.
Next, we evaluate the discrete latents on the ZeroSpeech ABX task to evaluate if the quantized features are able to discover phonetic units.
Our results show that representations learned by vq-wav2vec outperform vq-vae on both tasks. 
Context prediction is therefore a better task to learn discrete latent speech representations compared to autoencoding.

\section{Approaches}

In this section, we present the two different methods and their architectures for unsupervised speech pre-training.
The first approach is trained by reconstructing the input through a latent bottleneck between an encoder network and a decoder network~\cite{vqvae_wavenet2019} where the latent features serve as discrete representation.
The second approach learns through predicting the representations of future time-steps, which tasks the network to correctly identify true future time-steps from distractors, both of which are represented by discrete latent variables~\cite{baevski2019vqwav2vec}.

\begin{figure}[!htb]
    \centering
    \includegraphics[width=.55\textwidth]{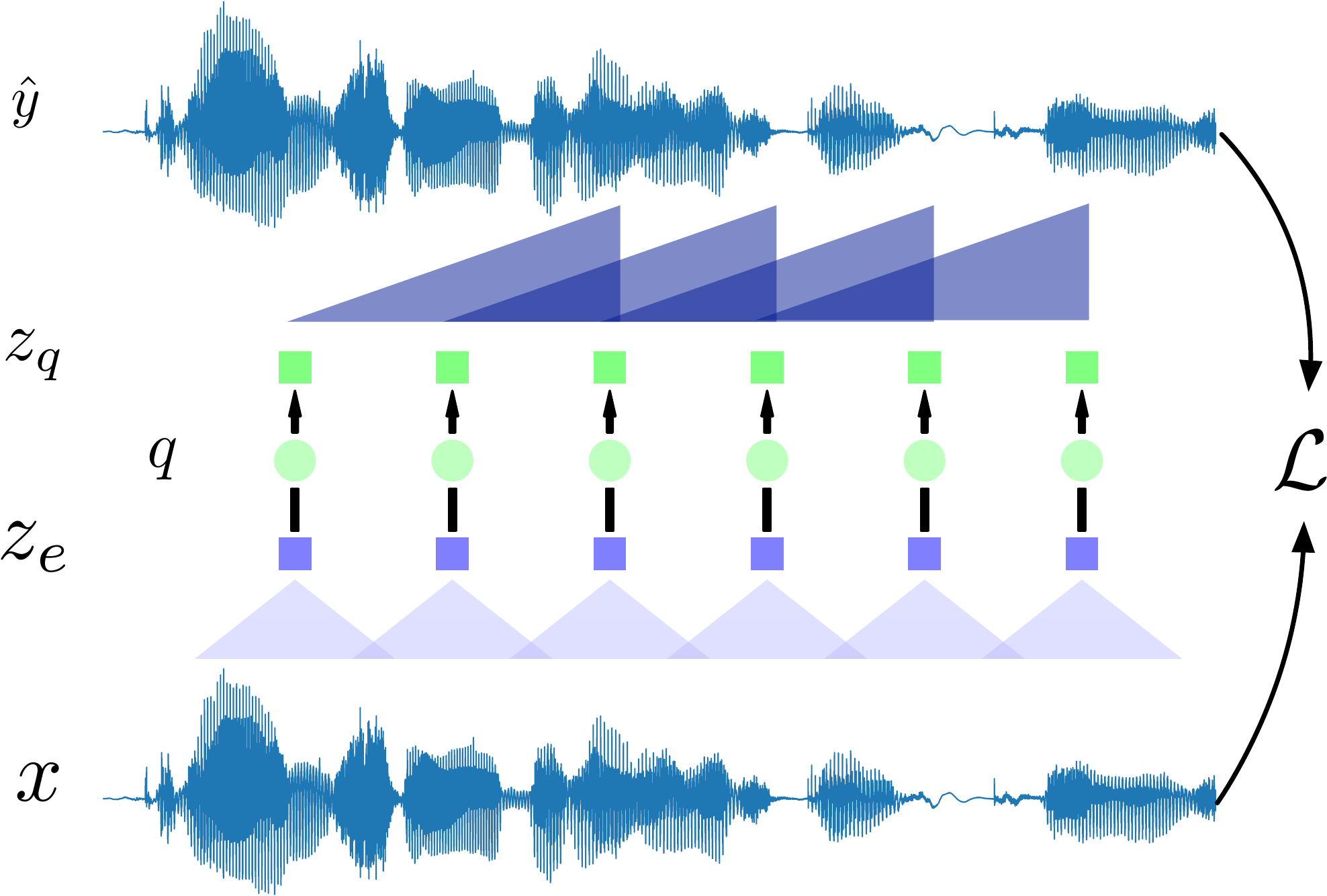}
    \caption{
            The vq-vae model is composed of an encoder which extracts dense features $z_e$ from the audio,
            the quantizer $q$ maps the dense features to discrete features $z_q$ and
            the decoder $p(x_t|x_{1...t-1}, z_q)$ reconstructs the original audio.
        }
    \label{fig:vqvae_diagram}
\end{figure}

\subsection{Autoencoding with vq-vae}

vq-vae~\cite{vqvae_wavenet2019} learns vector quantized representations by reconstructing the input. 
The audio data is first encoded as a sequence of dense representations $z_e$ which is then vector quantized through online k-means clustering (see~\cref{sec:vq_layer}) to obtain discrete vectors $z_q$.
Finally, an autoregressive decoder reconstructs the original waveform conditioned on past audio samples, the latent representation $z_q$, and optionally the speaker identity~\cite{vqvae_wavenet2019} (see Figure~\ref{fig:vqvae_diagram}).
To make the reconstruction feasible, the waveform is quantized to 256 levels through a mu-law transform~\cite{wavenet2016}.
The loss function for training vq-vae is
\begin{equation}\label{eqn:vqvae_loss}
    \mathcal{L}_{vq-vae} = -\text{log}p(x | z_q(x)) + \mathcal{L}^{\text{k-means}}
\end{equation}
The first term is the reconstruction loss of the audio in which we minimize a negative log-likelihood.
The second term denotes the loss for training the quantizer (\cref{sec:vq_layer}).

\subsection{Context-prediction with vq-wav2vec}

\begin{figure}[!htb]
    \centering
    \includegraphics[width=.55\textwidth]{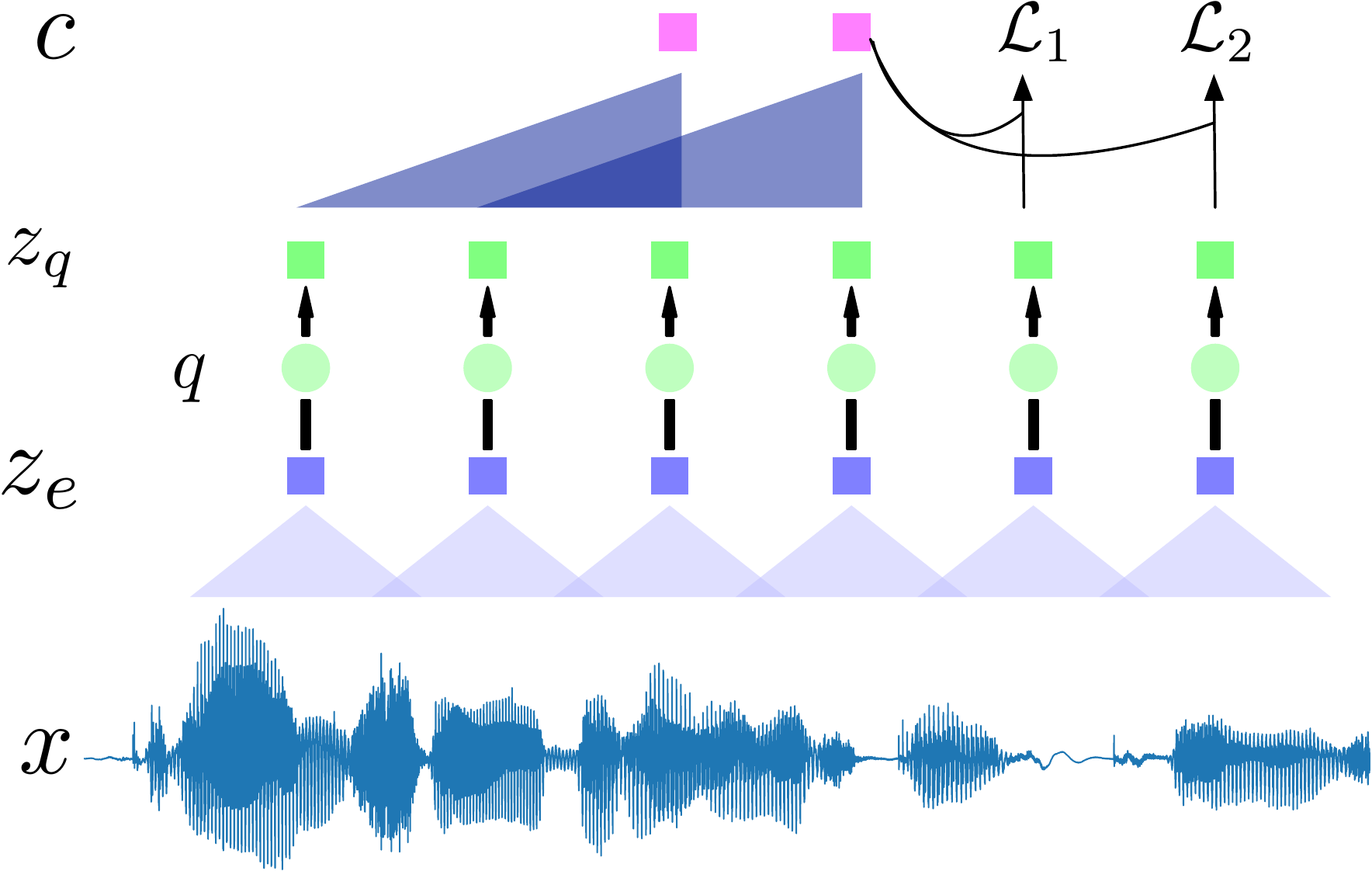}
    \caption{
        vq-wav2vec uses an encoder and a quantizer to compute dense and discrete features $z_e$ and $z_q$, respectively.
        An aggregator combines past discrete features into $c$.
        The loss is based on predicting future discrete latent speech representations based on the context features.}
    \label{fig:vqw2v_diagram}
\end{figure}

The vq-wav2vec model \cite{Baevski2020vq_wav2vec} learns the quantized vectors through predicting the representations of future timesteps. 
It consists of two networks applied to the raw audio signal (Figure~\ref{fig:vqw2v_diagram}).
An encoder network extracts the dense features ($z_e$) from the audio signal and a quantization module maps it to discrete features $z_q$.  
A context network then combines sequences of discrete features to obtain contextualized representations ($c_i$).
The model is trained to distinguish a future dense sample ${z_q}_{i+k}$ drawn from a set of distractors $p_n$ that are within $K$ time steps (\ref{fig:vqw2v_diagram}).
\begin{equation}
    \begin{aligned}
    \mathcal{L}^{\text{wav2vec}} 
    = - \displaystyle\sum_{i=1}^{T-k}(\text{log}\sigma({{z_e}_{i+k}}^\top h_k(c_i)) \\
                                 + \lambda \displaystyle \mathop{\mathbb{E}}_{ \Tilde{z}  \sim p_n}[ \sigma(-\Tilde{z}^\top h_k (c_i)) ])  
    \end{aligned}
\end{equation}
$T$ is the sequence length, $\sigma(x)=1/(1 + \exp{(-x)})$, $\sigma({{z_e}_{i+k}}^\top h_k(c_i))$ computes the probability of ${z_e}_{i+k}$ is the true sample.

\subsection{Vector quantization} \label{sec:vq_layer}

The above methods use either k-means \cite{oord2017vqvae} or Gumbel-Softmax \cite{jang2016categorical} to quantize high dimensional dense vectors.
A vector quantization layer maps a dense representation $z_e$ to a discrete representation $z_q$ from a codebook $\{e_i \in \mathbb{R}^D, i = [1...K]\}$ with $K$ entries.
\newline

\subsubsection{K-means}

K-means chooses the vector in the codebook which has the smallest Euclidean distance with respect to the input vector.
During training the loss is augmented by the following
\begin{equation}\label{eqn:kmeans_loss}
    \mathcal{L}^{\text{k-means}}
    = (z_e - [z_q])^2 + \beta ([z_e] - z_q)^2 
\end{equation}
where $[x] \equiv x$, $\frac{d}{dx}[x] \equiv 0$ is the stop-gradient operator.
During training, the gradient propagates through the dense vector $z_e$, while the selected vector $z_q$ from the codebook is updated by the Euclidean distance with respect to $z_e$.
\newline

\subsubsection{Gumbel-Softmax}

The Gumbel-Softmax~\cite{jang2016categorical} version of vq-wav2vec hard-selects a codeword $ z_q \in \mathbb{R}^D $ based on a linear projection $l \in \mathbb{R}^K $ and is fully differentiable.
The probability for selecting the $j$-th codeword is
\begin{equation}\label{eqn:vqw2v_gumbel_loss}
    p_j = \frac{\text{exp}(l_j + v_j) / \tau}{ \sum^{K}_{k=1} \text{exp}(l_k + v_k) / \tau }
\end{equation}
in which $v= -\text{log}(-\text{log}(u))$ where $u \sim \textit{U}(0, 1)$ and $\tau$ is the Gumbel softmax temperature.
During a forward pass, the hard index $\text{argmax}_jp_j$ is selected, and during the backward pass, the true gradient with respect to the logits is used.
\newline

\subsubsection{Codebook}
To avoid the problem of mode collapse of discrete latent models \cite{fastdecodeseqmodel2018}, we use several codebooks or groups~\cite{product_quantization_jegou11}.
A codebook is parametrized by the following parameters: $D$ the \textit{dimension} of the vector, $G$ the number of \textit{groups} in a codebook, $K$ the number of codewords within a group.

The codebook $\{z_q^i \in \mathbb{R} ^ {D / G} | i \in [1..M] \}$, where $M = K * G$ can represent $K^G$ distinct vectors of dimension $D$, and has a total number of $K \cdot D$ parameters.
We follow~\cite{Baevski2020vq_wav2vec} and share codewords across groups. 
In this way, the effective codebook can be represented as a matrix of shape $K \times (D / G)$.

\begin{figure}[!htb]
    \centering
    \includegraphics[width=.55\textwidth]{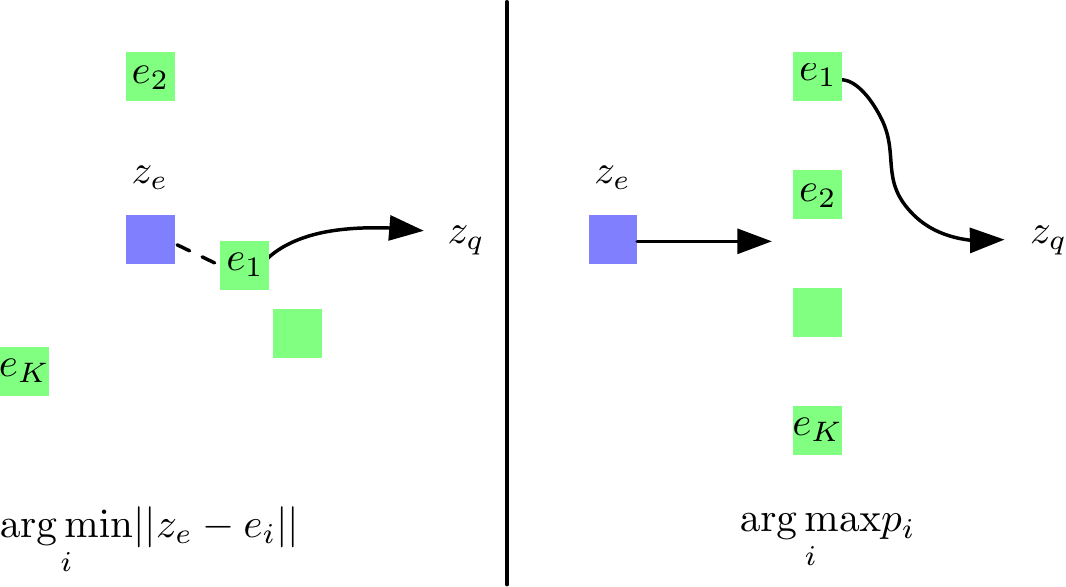}
    \caption{
            Two different methods for vector quantization.
            Left: K-Means selects a discretized latent variable based on the Euclidean distance to the dense representation $z_e$ in the vector space.
            Both vq-vae and the vq-wav2vec k-means version use this method.
            Right: Gumbel-Softmax based approach.
            $z_e$ is projected onto a vector and the latent with the largest logit is output.
    }
    \label{fig:quantizer_diagram}
\end{figure}

In a forward pass during inference, a dense feature $z_e \in \mathbb{R}^d$ is first organized into $G$ number of groups into the matrix form ${z_e}' \in \mathbb{R}^{G \times (d / G)}$.
Each row $j$ of ${z_e}'$ will then be converted into an integer index $i_j \in [1..K]$ through either nearest distance (vq-vae or vq-wav2vec Kmeans) $i_j = \argmin\limits_{k \in K} || {z_e}'_i - e_k ||^2 $ or largest value (vq-wav2vec Gumbel) $i_j = \argmax\limits_{k} {z_e}'_k$ (see Figure~\ref{fig:quantizer_diagram}).

During training with the Gumbel-Softmax, the true probability is being used for backward propagation.
\newline

\subsubsection{Codebook usage penalty}
Similar to~\cite{dieleman2018music,baevski2020wav}, we apply a penalty to encourage good usage of the codebook.
The softmax distribution $p_{g}$ is encouraged to be uniform across $K$ codewords within the codebook.
The entropy loss aims to maximize the entropy of the average probability of selection across group $g \in \{1...G\}$ and codewords $k \in \{1...K\}$ denoted $\bar{p}_{g,k}$ within a training batch.
\begin{equation}\label{eqn:diversity_loss}
    \mathcal{L}_{\text{diversity}} = \frac{1}{GK} \sum^{G} -H(\bar{p}_g) = \frac{1}{GK} \sum^{G,K}_{g=1,k=1} \bar{p}_{g,k} \text{log}\bar{p}_{g,k}
\end{equation}

\section{Experimental Setup}

\subsection{Datasets}
Models are pre-trained on Librispeech~\cite{librispeech2018}, except otherwise mentioned.
For evaluation we consider
TIMIT \cite{timit1993}, a 5-hour dataset of phoneme and speaker labels, and
ZeroSpeech2019 \cite{zerospeech2019} which contains 20-hours of audio without alignment, text or labels for Unit Discovery dataset.
For TIMIT, we apply the standard evaluation protocol and collapse detailed phoneme labels to 39 classes.

\subsection{Encoder Architecture}
\label{sec:encoder}

All approaches use the fairseq implementation~\cite{ott2019fairseq} of the wav2vec encoder for raw audio feature extraction~\cite{schneider2019wav2vec}.
The encoder contains 8 layers with 512 channels each, with kernel sizes (10, 8, 4, 4, 4, 1, 1, 1) and strides (5, 4, 2, 2, 2, 1, 1, 1), yielding a total receptive field of about 30 ms.

\subsection{Training details}

For vq-vae, we follow the same training scheme as in \cite{vqvae_wavenet2019} to train the vq-vae model but use the encoder architecture of vq-wav2vec.
We train for 300k updates using a cosine learning rate schedule \cite{cosinelr_16} with an initial learning rate of $2 \times 10^{-4}$ which is annealed to $1 \times 10^{-6}$.
The batch size is 64 with each audio sample having length 32ms~\cite{vqvae_wavenet2019}.
We experiment with inputting the speaker id to the decoder in order to learn speaker-independent representations.

For vq-wav2vec, we also train for 300k updates, warmup the learning rate from $1 \times 10^{-7}$ to $1 \times 10^{-4}$ over 500 updates, and then anneal to $1 \times 10^{-5}$ with a cosine schedule. 
We use a smaller batch size of 20, since we use larger segments of 150k frames (about 9.3 seconds).
Unless otherwise mentioned we use $K=320$, $G=2$ which has been found to be effective in previous work~\cite{baevski2019vqwav2vec}.

To evaluate the quality of the latent reprsenetation for phoneme recognition, we use a wav2letter acoustic model~\cite{pratap2019w2l}, trained for 1000 epochs on 8 GPUs for TIMIT.
The acoustic model is a convolutional neural network fed with raw audio and is optimized with an auto segmentation criterion.

\section{Results}

\subsection{Comparison of methods}

We train vq-vae and vq-wav2vec either on all audio data of Librispeech (960h) or the training data of ZeroSpeech 2019 (20h;~\cite{zerospeech2019}).
In our setup, the $z_q$ of both models have the same receptive field.
For vq-vae we use the k-means quantization scheme of~\cite{oord2017vqvae} with a single codebook of 4096 entries and a latent frequency of 50hz (\cref{sec:encoder}) which we found to work best on the validation set. 
For vq-wav2vec we consider both Gumbel ($G=8$, $K=8$) and k-means ($G=2$, $K=320$) with latent frequency of 50hz.

We evaluate the discrete representations output by the quantization modules of each model in several setups:
For TIMIT phoneme recognition we extract discrete features from the quantizer of vq-vae and vq-wav2vec and feed them into the acoustic model.
For the ABX task of ZeroSpeech 2019 we consider models trained on all of Librispeech (ZS(LS)) or the much smaller training data provided by the task (ZS).
The task evaluates whether the representations learned capture acoustic units.
We average all representations for a given sequence and then use a cosine distance to measure similarity between representations.

Table~\ref{tbl:main-results} shows that learning latent discrete representations of speech with context prediction (vq-wav2vec) performs generally better than autoencoding (vq-vae).
Reasoning about temporal information appears to result in better representations than reconstructing the raw audio.
Self-supervised work in natural language processing also heavily relies on objectives that require predicting future information or missing parts of the sequence~\cite{devlin2018bert,baevski2019clozedriven}.

For vq-vae, we do not see a large effect of using speaker ID.
We also compare to the recently introduced vq-cpc~\cite{niekerk2020vectorquantized}.
Similar to vq-wav2vec, vq-cpc combines vq-vae and cpc~\cite{oord2018cpc}.
However, different to vq-wav2vec, they input log-mel filterbanks instead of raw speech, they also use a much larger receptive field and use an RNN decoder instead of a WaveNet decoder. 
Finally, they perform model selection based on performance on the ZeroSpeech test set while as we selected models based on TIMIT validation performance.
vq-wav2vec (k-means) performs slightly better than their result and achieves a 13.22 error rate on the ABX task.

\begin{table}[t]
\centering
\caption{%
Comparison of vq-wav2vec with different quantizers as well as vq-vae (with k-means quantier) with and without speaker ID input to the decoder. We show phoneme error rate on the TIMIT test set, the error rate for the ZeroSpeech ABX evaluation (ZS) when training on the ZS training set as well as when training on Librispeech (ZS(LS)).
}
\label{tbl:main-results}
\begin{tabular}{lrrrr}
\toprule
{} & TIMIT & ZS(LS) & ZS  \\
\midrule
vq-cpc~\cite{niekerk2020vectorquantized} & - & - & 13.4 \\
\midrule
vq-wav2vec (Gumbel) & 16.54 & 14.12 & 15.37\\
vq-wav2vec (k-means) & 17.64 & 12.72 & 13.22\\
\midrule
vq-vae (w/ speaker)  & 19.99 & 18.61 & 18.73\\
vq-vae (w/o speaker) & 19.34 & 18.45 & 19.29\\
\bottomrule
\end{tabular}
\end{table}

\subsection{Codebook architecture of the quantizer}

In Table~\ref{tbl:codebook-results} we show that both models are sensitive to the choice of codebook architecture. 
In general, vq-wav2vec performs better when using multiple groups ($G$) while as vq-vae performs best when using a single codebook ($G=1$).

\begin{table}[t]
\centering
\caption{
ABX performanace on ZS(LS), cf. Table~\ref{tbl:main-results}, for a different number of entries in each codebook ($K$) and the number of codebooks or groups ($G$).
}
\label{tbl:codebook-results}
\begin{tabular}{lrr}
\toprule
{} & Codebook $(K \times G)$ & ZS(LS)  \\
\midrule
\multirow{4}{*}{\shortstack[l]{vq-wav2vec\\(Gumbel)}} & 4 x 8 & 14.2 \\
& 8 x 8 & 14.12 \\
& 320 x 2 & 14.18\\
& 512 x 1 & 14.77\\
\midrule
\multirow{4}{*}{\shortstack[l]{vq-wav2vec\\(k-means)}} & 4 x 8 & 15.06 \\
& 8 x 8 & 13.71 \\
& 320 x 2 & 12.72\\
& 512 x 1 & 18.41 \\
\midrule
\multirow{4}{*}{\shortstack[l]{vq-vae\\w/ speaker}} & 4 x 8 & 21.97 \\
& 8 x 8 & 24.93 \\
& 512 x 1 & 18.45 \\
& 320 x 2 & 19.59 \\
\bottomrule
\end{tabular}
\end{table}

\subsection{Analysis}

Next, we compute the co-occurence between human annotated phonemes and the discrete latent features produced by a vq-wav2vec model pretrained on the ZeroSpeech 20-hour data. 
We extract the discrete features $z_q$ on the TIMIT training data without any finetuning.
The TIMIT training data contains 3696 utterances of an average length 13.6 sec, equivalent to 563k discrete latents.

\begin{figure}[t]
\centering
\includegraphics[width=1.0\linewidth]{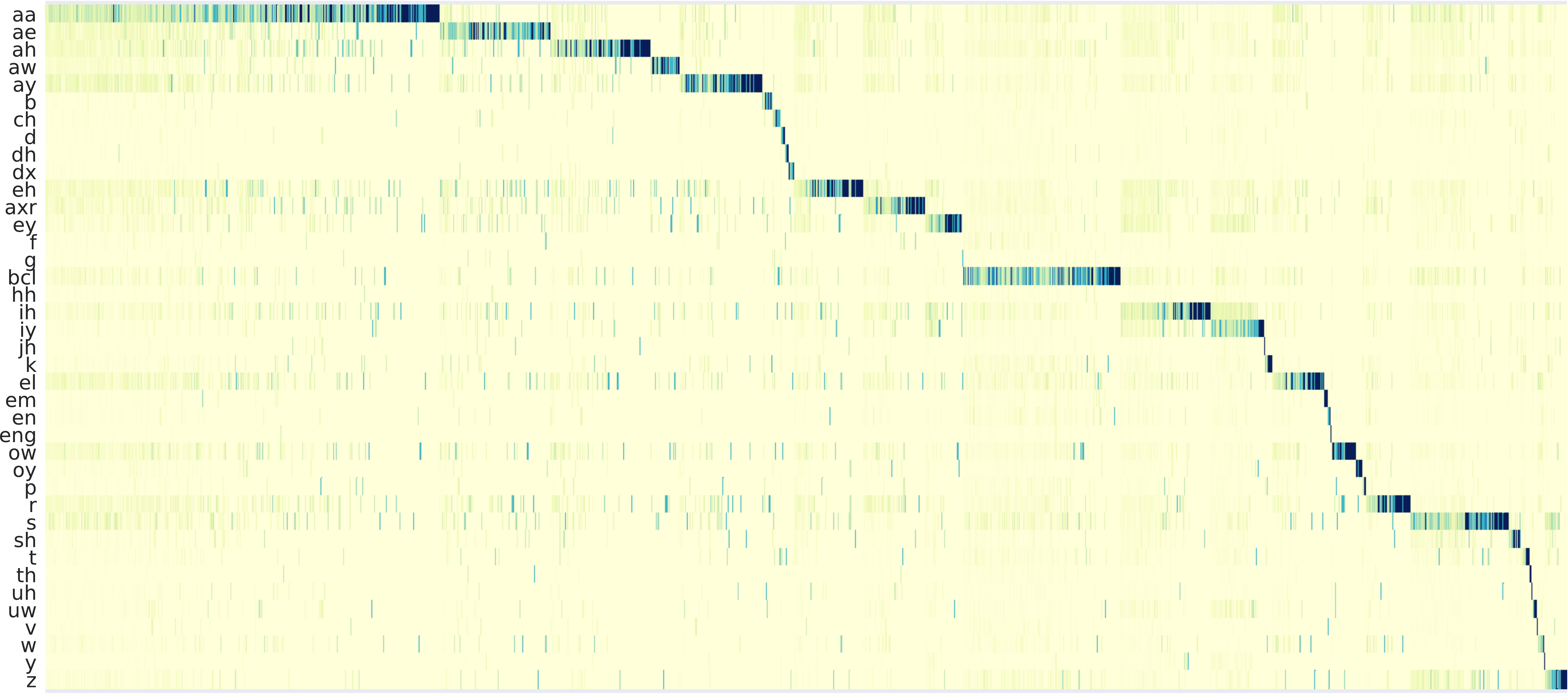}
\caption{
Visualization of the co-occurrence between discrete latent speech representations and phonemes.
We plot the conditional probability $P(phoneme|{z_q}_t)$ on TIMIT train data.
The y-axis shows the collapsed 39 classes of phonemes and the x-axis is over the different discrete latents.
}
\label{fig:token2phone_distribution}
\end{figure}

Figure~\ref{fig:token2phone_distribution} shows that many latents specialize in specific phonemes.
A large number of latents correspond to phonemes containing vowels, e.g., aa, ae, ah.
Similarly, there are many discrete latents which co-occur with silence (bcl) which is a frequent label in the TIMIT data.

\section{Conclusions}
We presented a comparison of two prevalent methods for learning discrete latent speech representations in terms of classical TIMIT phoneme recognition as well as the more recent ZeroSpeech ABX phoneme discrimination task.
Results show that predicting future time-steps is a more effective training task to learn these representations.
Future work includes exploring other objectives that require models to learn even richer representations.

\bibliography{refs}
\bibliographystyle{unsrtnat}

\end{document}